\documentclass[a4paper,twoside]{article}
\newcommand{\affil}[1]{$^{\rm #1}$}
\usepackage[authoryear]{natbib}
\bibpunct{(}{)}{;}{a}{}{,}
\usepackage{graphicx}
\usepackage{rotating}
\date{} %Please leave the date blank
\title{\large\bf\flushleft Pulsar Timing with the Parkes Radio Telescope for the {\it Fermi} Mission}
\author{\parbox{\textwidth}{\flushleft
\vspace{-0.5cm}
{\it P. Weltevrede\affil{A}, S. Johnston\affil{A,F},
R. N. Manchester\affil{A},
R. Bhat\affil{B},
M. Burgay\affil{C},
D. Champion\affil{A},
G. B. Hobbs\affil{A},
B. K{\i}z{\i}ltan\affil{D},
M. Keith\affil{A},
A. Possenti\affil{C},
J. E. Reynolds\affil{A},
K. Watters\affil{E}
}\\
\vspace{0.4cm}
{\small \affil{A}\,Australia Telescope National Facility, CSIRO, PO Box 76, Epping, NSW 1710}\\
{\small \affil{B}\,Centre for Astrophysics and Supercomputing, Swinburne University of Technology, Mail H39, PO Box 218, Hawthorn, Vic 3122}\\
{\small \affil{C}\,INAF-Osservatorio Astronomico di Cagliari, I-09012 Capoterra, Italy}\\
{\small \affil{D}\,Dept of Astronomy \& Astrophysics, University of California \& UCO Lick Observatory, Santa Cruz, CA 95064, USA}\\
{\small \affil{E}\,Dept of Physics, Stanford University, Stanford, CA 94305, USA}\\
{\small \affil{F}\,Email: Simon.Johnston@csiro.au}}}
\begin{document}
\twocolumn[
\begin{changemargin}{.8cm}{.5cm}
\begin{minipage}{.9\textwidth}
\vspace{-1cm}
\maketitle
%
%
%%%%%%%%%%%%%     ABSTRACT    %%%%%%%%%%%%%
%Abstract of no more than 200 words here.
\small{\bf Abstract:}
We report here on two years of timing of 168 pulsars using the Parkes
radio telescope. The vast majority of these pulsars have spin-down
luminosities in excess of 10$^{34}$~erg~s$^{-1}$ and are prime target
candidates to be detected in gamma-rays by the {\it Fermi} Gamma-Ray Space Telescope.
We provide the ephemerides for the ten pulsars being timed at Parkes 
which have been detected by {\it Fermi} in its first year of operation.
These ephemerides, in conjunction with the publicly available photon list,
can be used to generate gamma-ray profiles from the {\it Fermi} archive.
We will make the ephemerides of any pulsars of interest available to the 
community upon request.  In addition to the timing ephemerides, we present 
the parameters for 14 glitches which have 
occurred in 13 pulsars, seven of which have no previously known glitch history.
The Parkes timing programme, in conjunction with {\it Fermi} observations,
is expected to continue for at least the next four years.

%%%%%%%%%%%%%     KEYWORDS    %%%%%%%%%%%%%
\medskip{\bf Keywords:} pulsars:general --- pulsars:ephemerides --- pulsars:glitches
% Please write all keywords in lower case. PASA uses the
% standard list of subject headings adopted by The Astrophysical Journal
% and available from http://www.journals.uchicago.edu/ApJ/keywords_text.html.
% Keywords are separated by em-dashes, i.e. ---

%%%%%%%%DO NOT EDIT%%%%%%%%%%%%
\medskip
\medskip
\end{minipage}
\end{changemargin}
]
\small
%%%%%%%%EDIT FROM HERE%%%%%%%%%%%%

\section{Introduction}
The {\it Fermi} Gamma-Ray Space Telescope, launched in 2008 June,
has the study of pulsars at
gamma-ray wavelengths as one of its key science projects.
{\it Fermi} follows in the footsteps of its predecessor, the Compton Gamma-Ray
Observatory (CGRO) which was active during the 1990s. CGRO detected
seven gamma-ray pulsars (including the radio-quiet pulsar Geminga) with
three other possible detections (including a millisecond pulsar).
CGRO relied on ephemerides for the pulsars it observed in order to 
correctly phase tag the received gamma-ray photons to produce a light
curve. Three of the pulsars detected by CGRO (PSRs B1706$-$44, B1055$-$52 and
B1509$-$58) used data obtained with the Parkes radio telescope
(Thompson et al. 1992, Fierro et al. 1993, Ulmer et al. 1993,
Johnston et al. 1995) \nocite{tab+92,fbb+93,umw+93,jml+95}.

The Large Area Telescope (LAT) on board {\it Fermi}
(Atwood et al. 2009)\nocite{aaa+09} has a field of view which is substantially 
bigger than that of the Energetic Gamma-Ray Experiment Telescope (EGRET)
aboard CGRO, has a sensitivity more than an order of magnitude greater,
far superior energy resolution 
and a positional accuracy measured in arcmin rather than degrees.
{\it Fermi} began sky-survey observations on 2008 August 11 and in survey 
mode observes the entire sky every three hours.
The characteristics of the photons
received by {\it Fermi} remained proprietary for the first year of operation
but full data release occurred in 2009 August.

In principle, therefore, {\it Fermi} can produce light curves for every
pulsar in the sky and update the profile on a daily basis. To produce
these profiles for all but a handful of the brightest pulsars, generally
requires an accurate pulsar ephemeris so that the incoming
photons can be correctly phase tagged. Prior to the launch of {\it Fermi},
the worldwide radio pulsar timing community formalised a comprehensive
pulsar monitoring campaign using the Parkes, Lovell, Nan\c{c}ay, Green Bank
and Arecibo telescopes (Smith et al. 2008\nocite{sgc+08}).
The time pressure on these large
telescopes means that it is not possible to time all of the $\sim$2000
known pulsars. The timing list created by Smith et al. (2008) contained
224 pulsars with a spin-down energy loss rate, $\dot{E}$, larger than
10$^{34}$~erg~s$^{-1}$. Of these 224 pulsars, 156 are timed at Parkes.

One of the major issues facing the radio timing campaign is that many
of the pulsars potentially detectable by {\it Fermi} are young, high $\dot{E}$
objects. These pulsars are far from perfect clocks and suffer from
a high degree of timing noise \citep{hlk+04}. In order to track the
timing noise to a level of accuracy sufficient to time-tag the photons
to $10^{-2} - 10^{-3}$ of the rotational period, we need to observe the 
entire sample approximately once per month.

{\it Fermi} has been a spectacular success since its launch. It has discovered
16 gamma-ray pulsars through `blind' periodicity searches \citep{abdo5}
of which two have subsequently been detected in the radio
(Camilo et al. 2009)\nocite{cam09}.
In addition it has produced high quality profiles
for 8 millisecond radio pulsars \citep{abdo6}
and more than 20 `normal' (non-millisecond)
pulsars, including the Crab pulsar \citep{abdo8},
the Vela pulsar \citep{abdo3},
the recently discovered PSR~J1028--5819 \citep{kjk+08,abdo2}
and 6 pulsars with moderate $\dot{E}$ (Weltevrede et al. 2009)\nocite{welt09}.
All these results have been tabulated in the gamma-ray pulsar catalogue
\citep{abdo9}.

The radio data obtained through the pulsar timing program has a number of
other applications, including a statistical determination of the 
alignment of the 
magnetic and rotational axis as a function of time \citep{wj08a}
and examination of the polarisation characteristics of highly
energetic pulsars \citep{wj08b}.
In addition, long term radio timing shows that a number of pulsars glitch
\citep{sl96,wmp+00,js06} and the glitch parameters and time between events 
can be used to determine the interior structure of neutron stars \citep{rzc98}.

Of the `normal' radio pulsars detected by {\it Fermi}, Parkes is responsible 
for timing ten of them.  In this paper we describe the Parkes timing campaign
for the {\it Fermi} mission, including the observational 
and data analysis details.
We provide the timing ephemerides for, and a brief discussion of, the 10 
pulsars detected by {\it Fermi}. We tabulate the 
glitches seen in the monitored pulsars during the two years of the 
timing program.  We briefly discuss the implications of the results
and the next steps for the {\it Fermi} mission.

\section{Observations and Data Analysis}
Observations in support of the {\it Fermi} mission commenced in 2007 February.
All observations were carried out using the 64-m radio telescope in
Parkes, NSW, Australia.
Generally, each observing session used an observing frequency
near 1.4~GHz and lasted 24~hr with the observing sessions separated by 
approximately 4 weeks. However, once every six months we had an extended
observing session and data were obtained at both 3.1 and 0.7~GHz in
order to monitor long-term dispersion measure variations.
Each pulsar is typically observed for only a few minutes, sufficient
to reach a signal-to-noise ratio greater than 5.
A total of 156 pulsars from the Smith et al. (2008) list are timed at
Parkes; their basic parameters are given in Table~\ref{tab1a}.
A further 12 pulsars are also timed; these are listed separately in
Table~\ref{tab1c}.

At 1.4~GHz, we used the centre beam of the multibeam receiver \citep{swb+96}
with a total bandwidth of 256~MHz. The noise-equivalent flux density
of the system is $\sim$35~Jy on cold sky. For the 3.1~GHz and 0.7~GHz
observing, data were recorded simultaneously using the dual
10/50-cm receiver \citep{gzf+05} with a usable bandwidth of 1024 and 40~MHz
respectively. The noise-equivalent flux density is $\sim$49~Jy at 3.1~GHz
and 70~Jy at 0.7~GHz.  Many pulsars in our sample are
located in the Galactic plane where the sky temperature can be up
to $\sim$50~K at 1.4~GHz and several hundred K at 0.7~GHz, substantially 
raising the noise-equivalent flux density on these sources.

The downconverted signals from each polarisation channel of the linear
feeds were fed in a digital filterbank, designed and built at the
Australia Telescope National Facility.
The hardware converts the analogue voltages into digital signals
and produces a filterbank output consisting of 1024 frequency channels for
each of 1024 phase bins across the pulse period for the two auto correlations
and real and imaginary parts of the cross correlations of the feed probes.
Data were folded at the topocentric pulse period and accumulated for 
30~s and then dumped to disk.  A calibration signal, injected into the feed
at an angle of 45$^\circ$ to the probes, was recorded on a regular basis.
This signal is used to determine the relative gain of the two 
polarisation channels and the phase between them.

Data analysis was carried out using the {\sc psrchive} package \citep{hvm04}.
In brief, after removal of interference in both the frequency and
time domains, the data are gain and polarisation calibrated and
summed in frequency and time to produce a pulse profile for each
pulsar observed.

\section{Pulsar Timing}
In order to produce a time-of-arrival (TOA) the resultant pulse
profile is cross correlated with a high signal-to-noise ratio
`standard' profile. The standard profile is created by summing together
all previous observations of the pulsar using a technique described in
detail in \citet{wj08b}. The topocentric TOA at the weighted centre
frequency of the observation is then added to the database of previous
TOAs obtained for this pulsar.

Pulsar timing is carried out using the {\sc tempo2} package \citep{hem06}.
In addition to other auxiliary files, 
{\sc tempo2} requires a model for the spin behaviour of the pulsar
and the database of TOAs. After converting the TOAs to the solar system
barycentre using the DE405 model \citep{ehm06,sta98c},
{\sc tempo2} compares the TOAs to
the model predictions to produce a set of residuals.
These residuals are then used to refine the initial input model.

Assuming we have an accurate pulsar position obtained through other means,
the first step to determining a timing solution is to fit only for
the pulsar spin frequency, $\nu$, and frequency derivative, $\dot{\nu}$.
This effectively removes linear and quadratic terms from the residuals;
what remains is then generally dominated by a cubic term.
In these young pulsars, this cubic term is generally many orders of 
magnitude larger than the frequency second derivative expected from a 
simple dipole braking model and is the result of timing noise or glitch 
recovery (e.g. Johnston \& Galloway 1999). \nocite{jg99}
Although one can,
in principle, continue to fit higher order polynomials to the data in order to
whiten them, a different technique is used.
\begin{table*}[t]
\begin{center}
\caption{Parameters and post-fit rms in milli-periods (mP) for the pulsars 
taken from Smith et al. (2008) observed at Parkes in support of 
the {\it Fermi} mission.}
\label{tab1a}
\begin{tabular}{lrrcr || lrrcr}
\hline
Name & \multicolumn{1}{c}{P} & \multicolumn{1}{c}{DM} & log[$\dot{E}$] & rms &
Name & \multicolumn{1}{c}{P} & \multicolumn{1}{c}{DM} & log[$\dot{E}$] & rms \\
%\hline
%& (ms) & (cm$^{--3}$pc) & (ergs$^{-1}$) & (mP) &
%& (ms) & (cm$^{--3}$pc) & (ergs$^{-1}$) & (mP) \\
& \multicolumn{1}{c}{(ms)} & (pc/cm$^{3}$) & (erg/s) & (mP) &
& \multicolumn{1}{c}{(ms)} & (pc/cm$^{3}$) & (erg/s) & (mP) \\
\hline 
 J0543+2329 & 246.0 & 78 & 34.6 & 
1.1 & 
 J1302--6350 & 47.8 & 147 & 35.9 & 
2.8 \\ 
 J0614+2229 & 335.0 & 97 & 34.8 & 
3.3 & 
 J1305--6203 & 427.8 & 470 & 34.2 & 
2.0 \\ 
 J0627+0705 & 475.9 & 138 & 34.0 & 
0.4 & 
 J1320--5359 & 279.7 & 98 & 34.2 & 
0.5 \\ 
 J0659+1414 & 384.9 & 14 & 34.6 & 
3.2 & 
 J1327--6400 & 280.7 & 681 & 34.7 & 
8.1 \\ 
 J0729--1448 & 251.7 & 92 & 35.4 & 
5.3 & 
 J1341--6220 & 193.3 & 717 & 36.1 & 
3.7 \\ 
 J0742--2822 & 166.8 & 74 & 35.1 & 
0.9 & 
 J1349--6130 & 259.4 & 285 & 34.1 & 
0.9 \\ 
 J0745--5353 & 214.8 & 122 & 34.0 & 
4.2 & 
 J1357--6429 & 166.1 & 128 & 36.5 & 
7.0 \\ 
 J0821--3824 & 124.8 & 196 & 34.7 & 
1.9 & 
 J1359--6038 & 127.5 & 294 & 35.1 & 
0.2 \\ 
 J0834--4159 & 121.1 & 240 & 35.0 & 
3.9 & 
 J1406--6121 & 213.1 & 542 & 35.3 & 
9.4 \\ 
 J0835--4510 & 89.4 & 68 & 36.8 & 
2.3 & 
 J1412--6145 & 315.2 & 515 & 35.1 & 
3.6 \\ 
 J0855--4644 & 64.7 & 238 & 36.0 & 
12.0 & 
 J1413--6141 & 285.6 & 677 & 35.7 & 
89.7 \\ 
 J0857--4424 & 326.8 & 184 & 34.4 & 
1.4 & 
 J1420--6048 & 68.2 & 360 & 37.0 & 
16.7 \\ 
 J0901--4624 & 442.0 & 199 & 34.6 & 
0.6 & 
 J1452--5851 & 386.6 & 262 & 34.5 & 
2.2 \\ 
 J0905--5127 & 346.3 & 196 & 34.4 & 
0.4 & 
 J1452--6036 & 155.0 & 350 & 34.2 & 
0.4 \\ 
 J0908--4913 & 106.8 & 180 & 35.7 & 
0.5 & 
 J1453--6413 & 179.5 & 71 & 34.3 & 
0.6 \\ 
 J0940--5428 & 87.5 & 134 & 36.3 & 
4.5 & 
 J1509--5850 & 88.9 & 138 & 35.7 & 
12.1 \\ 
 J0954--5430 & 472.8 & 200 & 34.2 & 
0.7 & 
 J1512--5759 & 128.7 & 629 & 35.1 & 
0.9 \\ 
 J1003--4747 & 307.1 & 98 & 34.5 & 
0.5 & 
 J1513--5908 & 150.7 & 252 & 37.3 & 
11.2 \\ 
 J1015--5719 & 139.9 & 279 & 35.9 & 
4.8 & 
 J1514--5925 & 148.8 & 194 & 34.5 & 
3.2 \\ 
 J1016--5819 & 87.8 & 252 & 34.6 & 
2.0 & 
 J1515--5720 & 286.6 & 482 & 34.0 & 
1.5 \\ 
 J1016--5857 & 107.4 & 394 & 36.4 & 
3.2 & 
 J1524--5625 & 78.2 & 153 & 36.5 & 
5.3 \\ 
 J1019--5749 & 162.5 & 1039 & 35.3 & 
8.0 & 
 J1524--5706 & 1116.0 & 833 & 34.0 & 
0.6 \\ 
 J1020--6026 & 140.5 & 445 & 35.0 & 
12.9 & 
 J1530--5327 & 279.0 & 50 & 33.9 & 
0.8 \\ 
 J1043--6116 & 288.6 & 449 & 34.2 & 
0.5 & 
 J1531--5610 & 84.2 & 111 & 36.0 & 
1.5 \\ 
 J1048--5832 & 123.7 & 129 & 36.3 & 
5.3 & 
 J1538--5551 & 104.7 & 603 & 35.0 & 
8.4 \\ 
 J1052--5954 & 180.6 & 491 & 35.1 & 
12.1 & 
 J1539--5626 & 243.4 & 176 & 34.1 & 
0.9 \\ 
 J1055--6032 & 99.7 & 633 & 36.0 & 
--- & 
 J1541--5535 & 295.8 & 428 & 35.0 & 
5.3 \\ 
 J1057--5226 & 197.1 & 30 & 34.5 & 
0.3 & 
 J1543--5459 & 377.1 & 346 & 34.6 & 
1.9 \\ 
 J1105--6107 & 63.2 & 271 & 36.4 & 
5.2 & 
 J1548--5607 & 170.9 & 316 & 34.9 & 
1.8 \\ 
 J1112--6103 & 65.0 & 599 & 36.7 & 
10.5 & 
 J1549--4848 & 288.3 & 56 & 34.4 & 
1.2 \\ 
 J1115--6052 & 259.8 & 228 & 34.2 & 
0.7 & 
 J1551--5310 & 453.4 & 493 & 34.9 & 
7.1 \\ 
 J1119--6127 & 407.7 & 707 & 36.4 & 
0.2 & 
 J1600--5044 & 192.6 & 261 & 34.4 & 
0.4 \\ 
 J1123--6259 & 271.4 & 223 & 34.0 & 
0.9 & 
 J1600--5751 & 194.5 & 177 & 34.0 & 
0.9 \\ 
 J1124--5916 & 135.3 & 330 & 37.1 & 
--- & 
 J1601--5335 & 288.5 & 195 & 35.0 & 
9.4 \\ 
 J1138--6207 & 117.6 & 520 & 35.5 & 
4.9 & 
 J1611--5209 & 182.5 & 128 & 34.5 & 
0.3 \\ 
 J1156--5707 & 288.4 & 244 & 34.6 & 
0.7 & 
 J1614--5048 & 231.7 & 583 & 36.2 & 
5.6 \\ 
 J1216--6223 & 374.0 & 787 & 34.1 & 
4.4 & 
 J1617--5055 & 69.4 & 467 & 37.2 & 
--- \\ 
 J1224--6407 & 216.5 & 97 & 34.3 & 
0.3 & 
 J1626--4807 & 293.9 & 817 & 34.4 & 
17.8 \\ 
 J1248--6344 & 198.3 & 433 & 34.9 & 
8.0 & 
 J1627--4706 & 140.7 & 456 & 34.4 & 
9.4 \\ 
 J1301--6305 & 184.5 & 374 & 36.2 & 
14.0 & 
 J1632--4757 & 228.6 & 578 & 34.7 & 
7.8 \\ 
\hline
\end{tabular}
\end{center}
\end{table*}

\begin{table*}[t]
\addtocounter{table}{-1}
\begin{center}
\caption{Parameters and post-fit rms in milli-periods (mP) for the pulsars 
taken from Smith et al. (2008) observed at Parkes in support of 
the {\it Fermi} mission (cont).}
\label{tab1b}
\begin{tabular}{lrrcr || lrrcr}
\hline
Name & \multicolumn{1}{c}{P} & \multicolumn{1}{c}{DM} & log[$\dot{E}$] & rms &
Name & \multicolumn{1}{c}{P} & \multicolumn{1}{c}{DM} & log[$\dot{E}$] & rms \\
%\hline
%& (ms) & (cm$^{--3}$pc) & (ergs$^{-1}$) & (mP) &
%& (ms) & (cm$^{--3}$pc) & (ergs$^{-1}$) & (mP) \\
& \multicolumn{1}{c}{(ms)} & (pc/cm$^{3}$) & (erg/s) & (mP) &
& \multicolumn{1}{c}{(ms)} & (pc/cm$^{3}$) & (erg/s) & (mP) \\
\hline 
 J1632--4818 & 813.5 & 758 & 34.7 & 
12.8 & 
 J1801--2451 & 124.9 & 289 & 36.4 & 
4.9 \\ 
 J1637--4553 & 118.8 & 193 & 34.9 & 
0.9 & 
 J1803--2137 & 133.6 & 234 & 36.3 & 
6.5 \\ 
 J1637--4642 & 154.0 & 417 & 35.8 & 
9.9 & 
 J1806--2125 & 481.8 & 750 & 34.6 & 
3.8 \\ 
 J1638--4417 & 117.8 & 436 & 34.6 & 
4.0 & 
 J1809--1917 & 82.7 & 197 & 36.3 & 
5.3 \\ 
 J1638--4608 & 278.1 & 424 & 35.0 & 
0.9 & 
 J1812--1910 & 431.0 & 892 & 34.3 & 
9.0 \\ 
 J1640--4715 & 517.4 & 592 & 34.1 & 
3.7 & 
 J1815--1738 & 198.4 & 728 & 35.6 & 
3.3 \\ 
 J1643--4505 & 237.4 & 484 & 35.0 & 
2.4 & 
 J1820--1529 & 333.2 & 772 & 34.6 & 
5.5 \\ 
 J1646--4346 & 231.6 & 490 & 35.6 & 
6.2 & 
 J1824--1945 & 189.3 & 225 & 34.5 & 
0.3 \\ 
 J1648--4611 & 165.0 & 393 & 35.3 & 
6.2 & 
 J1825--1446 & 279.2 & 357 & 34.6 & 
1.7 \\ 
 J1649--4653 & 557.0 & 332 & 34.0 & 
1.6 & 
 J1826--1334 & 101.5 & 231 & 36.4 & 
7.7 \\ 
 J1650--4502 & 380.9 & 320 & 34.0 & 
2.0 & 
 J1828--1057 & 246.3 & 245 & 34.7 & 
6.7 \\ 
 J1650--4921 & 156.4 & 230 & 34.3 & 
1.0 & 
 J1828--1101 & 72.1 & 607 & 36.2 & 
5.1 \\ 
 J1702--4128 & 182.1 & 367 & 35.5 & 
6.5 & 
 J1830--1059 & 405.0 & 162 & 34.6 & 
1.0 \\ 
 J1702--4305 & 215.5 & 538 & 34.6 & 
5.9 & 
 J1831--0952 & 67.3 & 247 & 36.0 & 
6.3 \\ 
 J1702--4310 & 240.5 & 377 & 35.8 & 
4.1 & 
 J1832--0827 & 647.3 & 301 & 34.0 & 
0.7 \\ 
 J1705--3950 & 318.9 & 207 & 34.9 & 
2.0 & 
 J1833--0827 & 85.3 & 411 & 35.8 & 
1.7 \\ 
 J1709--4429 & 102.5 & 76 & 36.5 & 
4.6 & 
 J1834--0731 & 513.0 & 295 & 34.2 & 
4.3 \\ 
 J1715--3903 & 278.5 & 313 & 34.8 & 
5.7 & 
 J1835--0643 & 305.8 & 473 & 34.7 & 
4.6 \\ 
 J1718--3825 & 74.7 & 247 & 36.1 & 
3.0 & 
 J1835--0944 & 145.3 & 277 & 34.7 & 
5.9 \\ 
 J1721--3532 & 280.4 & 496 & 34.7 & 
1.0 & 
 J1835--1106 & 165.9 & 133 & 35.3 & 
2.4 \\ 
 J1722--3712 & 236.2 & 100 & 34.5 & 
1.7 & 
 J1837--0559 & 201.1 & 318 & 34.2 & 
4.4 \\ 
 J1723--3659 & 202.7 & 254 & 34.6 & 
1.1 & 
 J1837--0604 & 96.3 & 462 & 36.3 & 
17.1 \\ 
 J1726--3530 & 1110.3 & 727 & 34.5 & 
6.3 & 
 J1838--0453 & 380.8 & 621 & 34.9 & 
2.8 \\ 
 J1730--3350 & 139.5 & 259 & 36.1 & 
1.3 & 
 J1838--0549 & 235.3 & 274 & 35.0 & 
4.3 \\ 
 J1731--4744 & 829.8 & 123 & 34.0 & 
0.4 & 
 J1839--0321 & 238.8 & 449 & 34.6 & 
4.1 \\ 
 J1733--3716 & 337.6 & 154 & 34.2 & 
1.0 & 
 J1839--0905 & 419.0 & 348 & 34.1 & 
1.9 \\ 
 J1734--3333 & 1169.2 & 578 & 34.7 & 
9.6 & 
 J1841--0425 & 186.1 & 325 & 34.6 & 
0.5 \\ 
 J1735--3258 & 351.0 & 754 & 34.4 & 
14.9 & 
 J1841--0524 & 445.7 & 289 & 35.0 & 
4.6 \\ 
 J1737--3137 & 450.4 & 488 & 34.8 & 
4.2 & 
 J1842--0905 & 344.6 & 343 & 34.0 & 
1.1 \\ 
 J1738--2955 & 443.4 & 223 & 34.6 & 
0.9 & 
 J1843--0355 & 132.3 & 798 & 34.3 & 
12.4 \\ 
 J1739--2903 & 322.9 & 139 & 34.0 & 
0.3 & 
 J1843--0702 & 191.6 & 228 & 34.1 & 
2.6 \\ 
 J1739--3023 & 114.4 & 170 & 35.5 & 
5.8 & 
 J1844--0256 & 273.0 & 820 & 34.7 & 
23.7 \\ 
 J1740--3015 & 606.8 & 152 & 34.9 & 
9.6 & 
 J1844--0538 & 255.7 & 413 & 34.4 & 
1.1 \\ 
 J1745--3040 & 367.4 & 88 & 33.9 & 
0.7 & 
 J1845--0743 & 104.7 & 281 & 34.1 & 
0.3 \\ 
 J1756--2225 & 405.0 & 326 & 34.5 & 
1.1 & 
 J1847--0402 & 597.8 & 142 & 34.0 & 
0.7 \\ 
 J1757--2421 & 234.1 & 179 & 34.6 & 
0.4 & 
 J1853+0011 & 397.9 & 569 & 34.3 & 
2.9 \\ 
 J1801--2154 & 375.3 & 388 & 34.1 & 
3.1 & 
 J1853--0004 & 101.4 & 438 & 35.3 & 
1.6 \\ 
 J1801--2304 & 415.8 & 1074 & 34.8 & 
4.0 & 
 J1903+0925 & 357.2 & 162 & 30.7 & 
--- \\ 
\hline
\end{tabular}
\end{center}
\end{table*}

The technique (developed by
Hobbs et al. 2004\nocite{hlk+04}) instead removes harmonically
related sinusoids from the residual data after fitting for 
$\nu$ and $\dot{\nu}$ and continues to remove components until the 
residuals resemble white noise. This whitening technique is implemented
using the {\sc fitwaves} algorithm within {\sc tempo2}.

An added complication arises if the pulsar suffers a glitch. A glitch causes an
abrupt change in both $\nu$ and $\dot{\nu}$ and the magnitude of the change
can be large. With sparsely sampled data it is very difficult to maintain 
phase coherence through the epoch of the glitch.
This implies that the glitch
epoch cannot be well determined -- in the case of our data, we generally
cannot determine the glitch epoch to better than $\sim$15 days.
In order to measure the glitch parameters, we
determine a timing solution before and after the glitch and then, 
using the two solutions, estimate $\Delta\nu$ and $\Delta\dot{\nu}$.
With these estimates, and by setting the glitch epoch midway between 
the observation just before and just after the glitch,
a timing solution can be obtained for the entire data set 
with the addition of a phase jump at the glitch epoch.

One of the important parameters to measure, in the context of models of 
gamma-ray emission, is the offset between the
radio and gamma-ray profiles. In order to do this, an accurate value
of the dispersion measure (DM) is needed so that the dispersion delay
between the $\sim$1~GHz radio profile and the $\sim$GeV gamma-ray profile
can be corrected for. We measured the DM across the 256~MHz bandwidth
at 1.4~GHz assuming no profile evolution across the band which is generally
sufficient to determine the DM to $\sim$0.01~cm$^{-3}$pc. This corresponds
to an uncertainty of $\sim$40~$\mu$s in the offset between the 
radio and gamma-ray profiles.
\begin{table}[t]
\begin{center}
\caption{Parameters and post-fit rms in milli-periods (mP) for the pulsars
observed in addition to those contained in Smith et al. (2008).}
\label{tab1c}
\begin{tabular}{lrrcr}
\hline
Name & \multicolumn{1}{c}{P} & \multicolumn{1}{c}{DM} & log[$\dot{E}$] & rms \\
& \multicolumn{1}{c}{(ms)} & (pc/cm$^{3}$) & (erg/s) & (mP) \\
\hline 
J0108--1431 & 807.6 & 2 & 30.8 & 
0.5 \\
J0401--7608 & 545.3 & 22 & 32.6 & 
1.0 \\
J0536--7543 & 1245.9 & 18 & 31.1 & 
0.4 \\
J0630--2834 & 1244.4 & 34 & 32.2 & 
1.2 \\
J0738--4042 & 374.9 & 161 & 33.1 & 
0.6 \\
J1028--5819 & 91.4 & 97 & 35.9 & 
0.4 \\
J1456--6843 & 263.4 & 9 & 32.3 & 
15.9 \\
J1602--5100 & 864.2 & 171 & 33.6 & 
0.9 \\
J1638--4725 & 763.9 & 550 & 32.4 & 
52.1 \\
J1705--1906 & 299 & 23 & 33.8 & 
0.3 \\
J1825--0935 & 769 & 19 & 33.7 & 
0.9 \\
J1845--0434 & 486.8 & 231 & 33.6 & 
0.8 \\
\hline
\end{tabular}
\end{center}
\end{table}

\begin{table}[t]
\begin{center}
\caption{Parameters of the 14 glitches observed in the timing data.}
\label{tabglitch}
\begin{tabular}{lccc}
\hline Name & Epoch & $\Delta{\nu}/{\nu}$ & $\Delta{\dot{\nu}}/\dot{\nu}$ \\
& (MJD) & ($\times 10^{-6}$) & ($\times 10^{-3}$) \\
\hline
J0729--1448$^\dag$ & 54711(21) & 6.6203(4) & 13.86(5)\\
J1048--5832 & 54495(10) & 3.0431(2) & 4.77(2)\\
J1052--5954$^\dag$ & 54526(22) & 6.7478(7) & 14.1(4)\\
J1105--6107 & 54711(21) & 0.023(1) & 0.0(3)\\
J1341--6220 & 54468(18) & 3.0782(3)\\
            & 54870(11) & 3.0667(3)\\
J1357--6429 & 54803(17) & 1.752(7) & 2.8(2.8)\\
J1410--6132$^\dag$ & 54652(19) & 0.2726(5) & --3.31(5)\\
J1420--6048$^\dag$ & 54652(20) & 0.9371(3) & 2.95(1)\\
J1709--4429 & 54710(22) & 2.7497(1) & 4.95(1)\\
J1737--3137$^\dag$ & 54352.3(1.0) & 1.3425(4) & 0.4(6)\\
J1740--3015$^\dag$ & 54468(17) & 0.037(1) & --0.4(4)\\
J1801--2451 & 54680(9) & 3.0904(2) & 5.04(2)\\
J1841--0524$^\dag$ & 54495(10) & 1.0359(4) & 0.11(15)\\
\hline
\end{tabular}
\medskip\\
$^\dag$Pulsar not previously known to have glitched.
\end{center}
\end{table}

\begin{figure*}[t]
\begin{center}
\begin{tabular}{cc}
\includegraphics[scale=0.3, angle=-90]{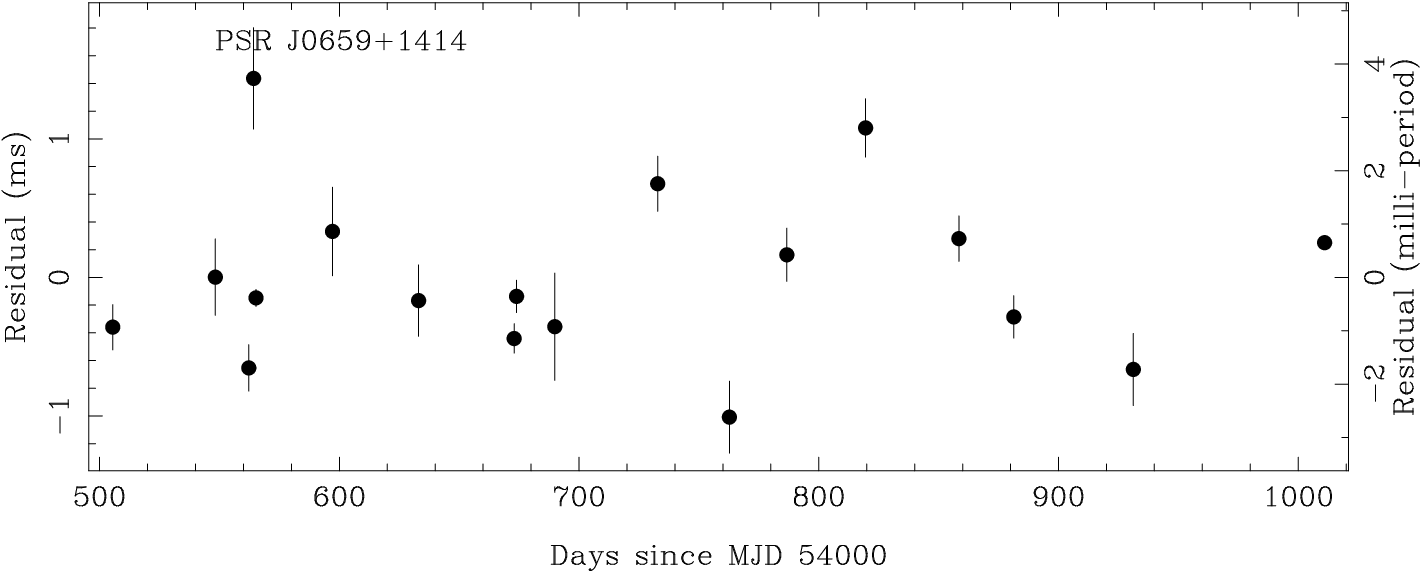} &
\includegraphics[scale=0.3, angle=-90]{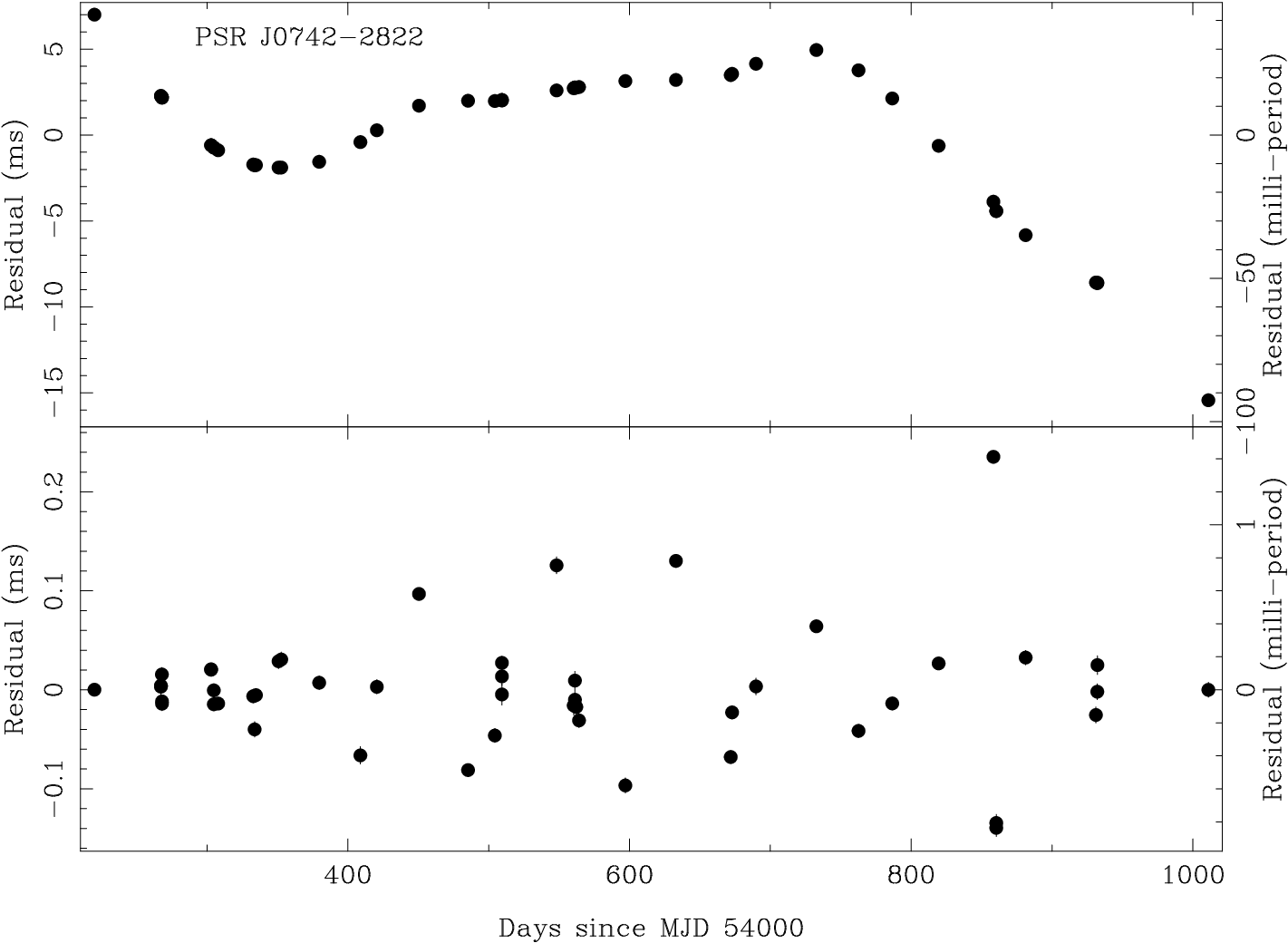} \\
\includegraphics[scale=0.3, angle=-90]{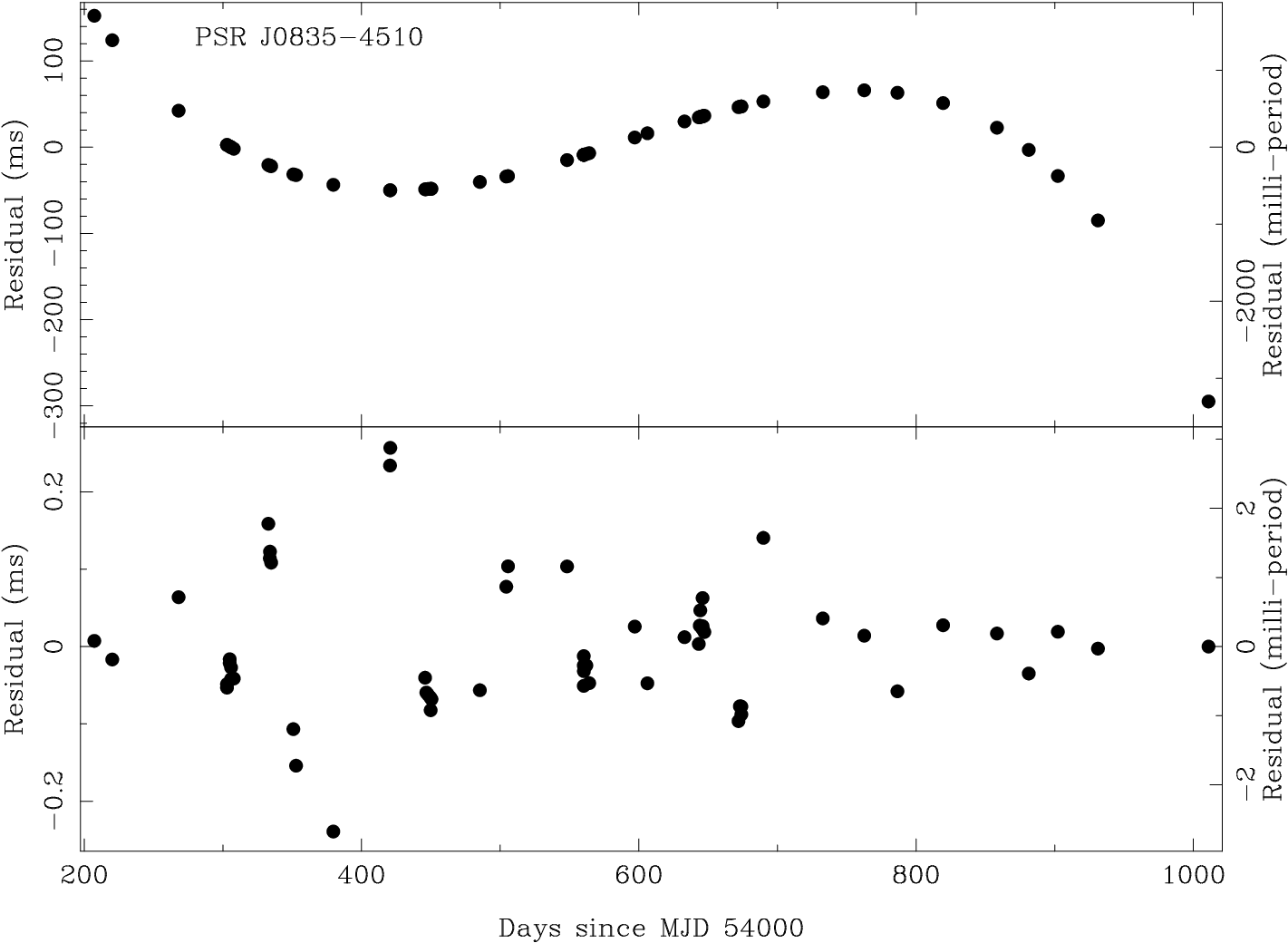} &
\includegraphics[scale=0.3, angle=-90]{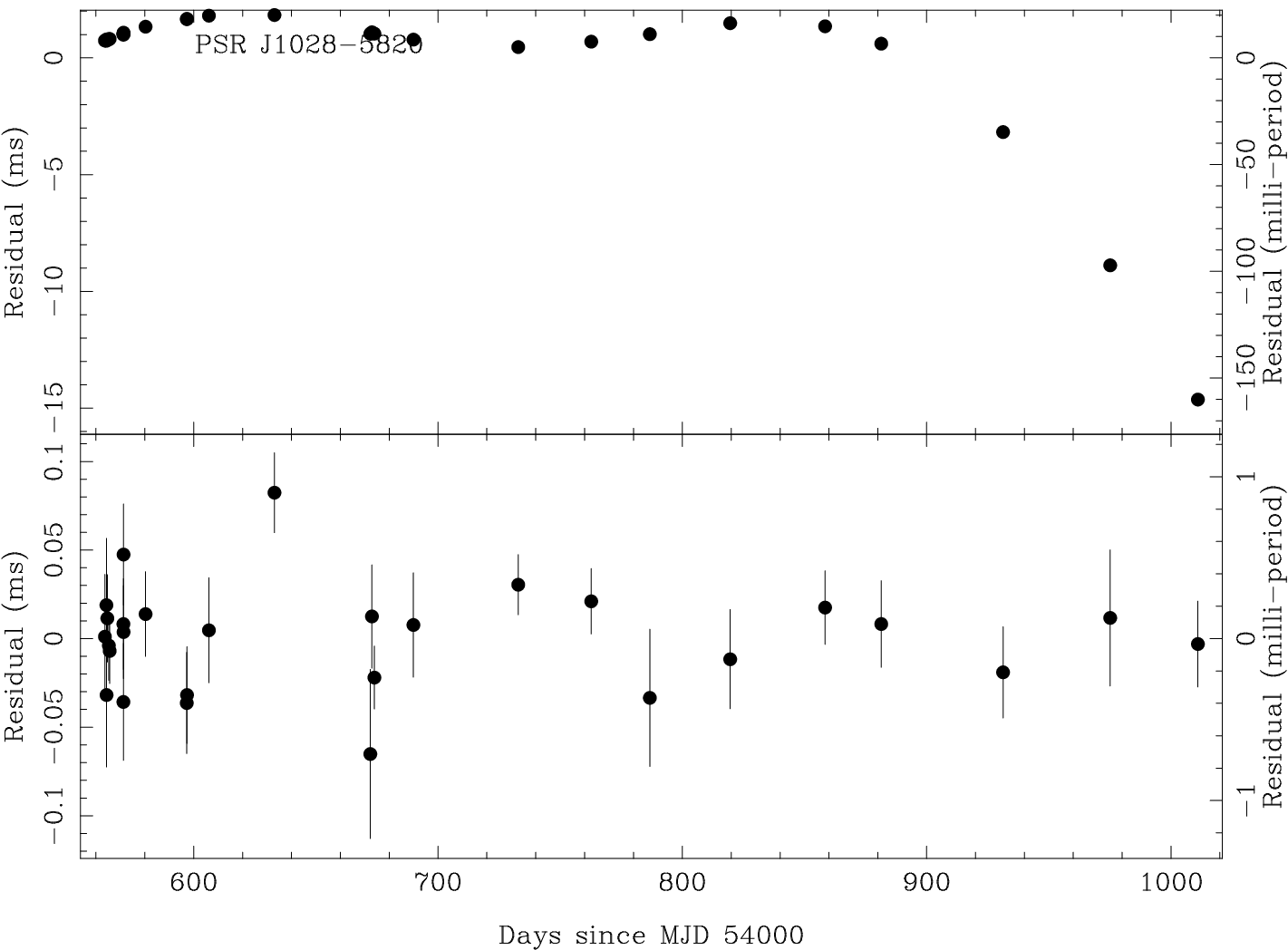} \\
\end{tabular}
\end{center}
\caption{Timing residuals for PSRs~J0659+1414 (top left),
J0742--2822 (top right), J0835--4510 (bottom left) and J1028--5819
(bottom right). Residuals before (top) and after (bottom) the use of
{\sc fitwaves} are shown for the latter three pulsars.
Note the very different scales on the y-axis for these cases.}
\label{figtime1}
\end{figure*}

\section{Results - General Timing}
In the last columns of Tables~\ref{tab1a} and \ref{tab1c}
we list the rms of the residuals in milli-periods for each pulsar following
fitting for $\nu$, $\dot{\nu}$ and whitening if necessary.
When creating gamma-ray profiles from the radio ephemeris, the rms
constrains the maximum useful number of phase bins across the
light curve. The rms is less than 2 milli-periods for 64 pulsars
in our sample -- 500 bins across the light curve is then possible.
Only for 18 pulsars is the rms worse than 10 milli-periods.
We note that the accuracy of the time tagging of the gamma-ray photons is
significantly better than 1~$\mu$s (Smith et al. 2008) and is not a limiting
factor in the time-resolution of the resulting gamma-ray profiles.
Typically, the small number of photons received from gamma-ray pulsars
limits the profile resolution to less than 100 bins.

In three cases we have been unable to maintain phase coherence in the
timing data. The pulsars concerned are PSRs J1055--6032, J1617--5055
and J1903+0925. In the latter case we have less than 1~yr of timing data
but the first two named pulsars appear to have extremely large
short-term timing noise.  They need to be monitored on a much more 
regular basis than once per month in order to maintain phase coherence.
In addition, PSR~J1124--5916 is an extremely weak 
radio pulsar \citep{cmg+02} for which it
takes $\sim$5~hr to obtain a timing point using Parkes and it is therefore
not observed as part of our timing programme. Following its detection
in gamma-rays by {\it Fermi} \citep{abdo9}, we observed it on a single
occasion in order to phase align the radio and gamma-ray profiles.

\section{Results - Glitches}
We have detected 14 glitches in 13 of the pulsars in our sample.
These glitches are tabulated in Table~\ref{tabglitch} with the parameters
estimated using the scheme outlined in Section~3.
The epoch of the glitch is set at the midpoint between 
the last pre-glitch and first post-glitch data point. We note that
{\it Fermi} data can be used to better constrain the glitch epoch for
PSR~J1709--4429 \citep{abdo11}.
The third column of the table gives the fractional change
in the spin frequency, $\Delta\nu/\nu$ and the final column lists
the fractional change in the frequency derivative, $\Delta\dot{\nu}/\dot{\nu}$.
The listed errors are twice the formal
errors given by {\sc tempo2} and given in brackets which refer to 
the last digit(s).
Three of the pulsars in the table have been detected by {\it Fermi}. For PSRs
J1048--5832 and J1420--6048 the glitch occurred prior to the launch of the 
satellite. The glitch in  PSR~J1709--4429 happened shortly after {\it Fermi} 
began its all sky survey in 2008 August.

Of the 13 pulsars, 7 have not previously been known to glitch, these are listed
with the $\dag$ symbol in Table~\ref{tabglitch}. The other 6 pulsars have
previously been observed to glitch as tabulated in the compilation of 
\citet{wmp+00}. Most notable is PSR~J1341--6220 (B1338--62) which has the
shortest average interval between glitches of any pulsar, estimated
by \citet{wmp+00} to be 250 days.
In the 800 days of our timing data we see two glitches
separated by some 400 days. Both glitches have about the same magnitude
and are larger than any of the other 12 glitches reported for this pulsar.
There is not sufficient time between the glitches to determine a
reliable value of $\Delta\dot{\nu}/\dot{\nu}$.
PSR~J1740--3015 (B1737--30) is another pulsar which glitches often
\citep{zwm+08} and we have detected one relatively small glitch in this object.

The smallest glitch we report here has $\Delta\nu/\nu$ of
$2\times 10^{-8}$ and we believe our sample to be complete
above a limit of about $10^{-8}$. Below this limit it is very
hard, with our data set, to distinguish between glitches and 
timing noise. We attempted to
fit glitch parameters to possible `cuspy' looking residuals,
similar to those seen in \citet{js06} for example,
but we find that fitting glitches and using {\sc fitwaves} return very 
similar results in spite of their different functional forms. 

\section{Results - {\it Fermi} pulsars}
\begin{figure*}[t]
\begin{center}
\begin{tabular}{cc}
\includegraphics[scale=0.3, angle=-90]{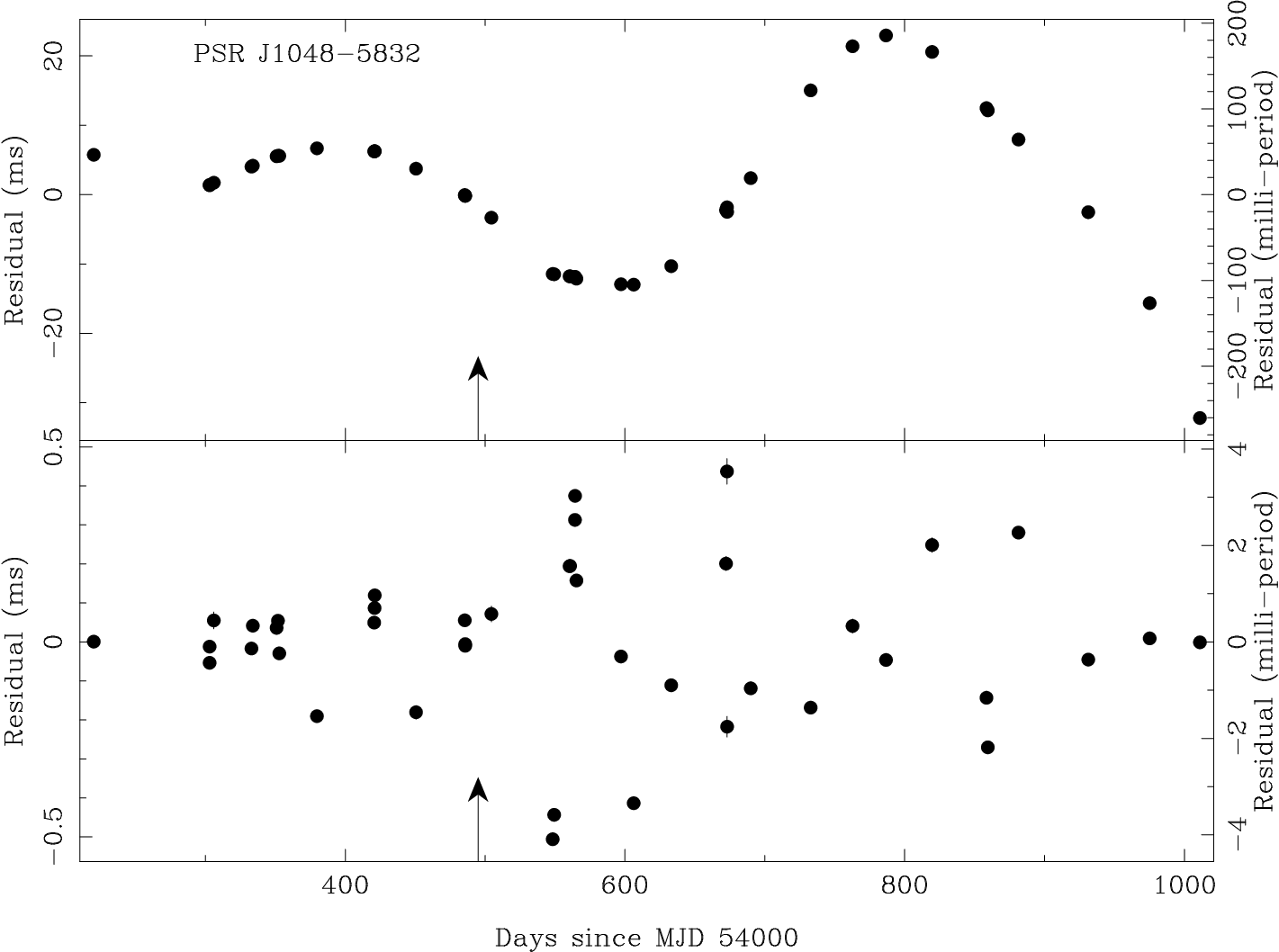} &
\includegraphics[scale=0.3, angle=-90]{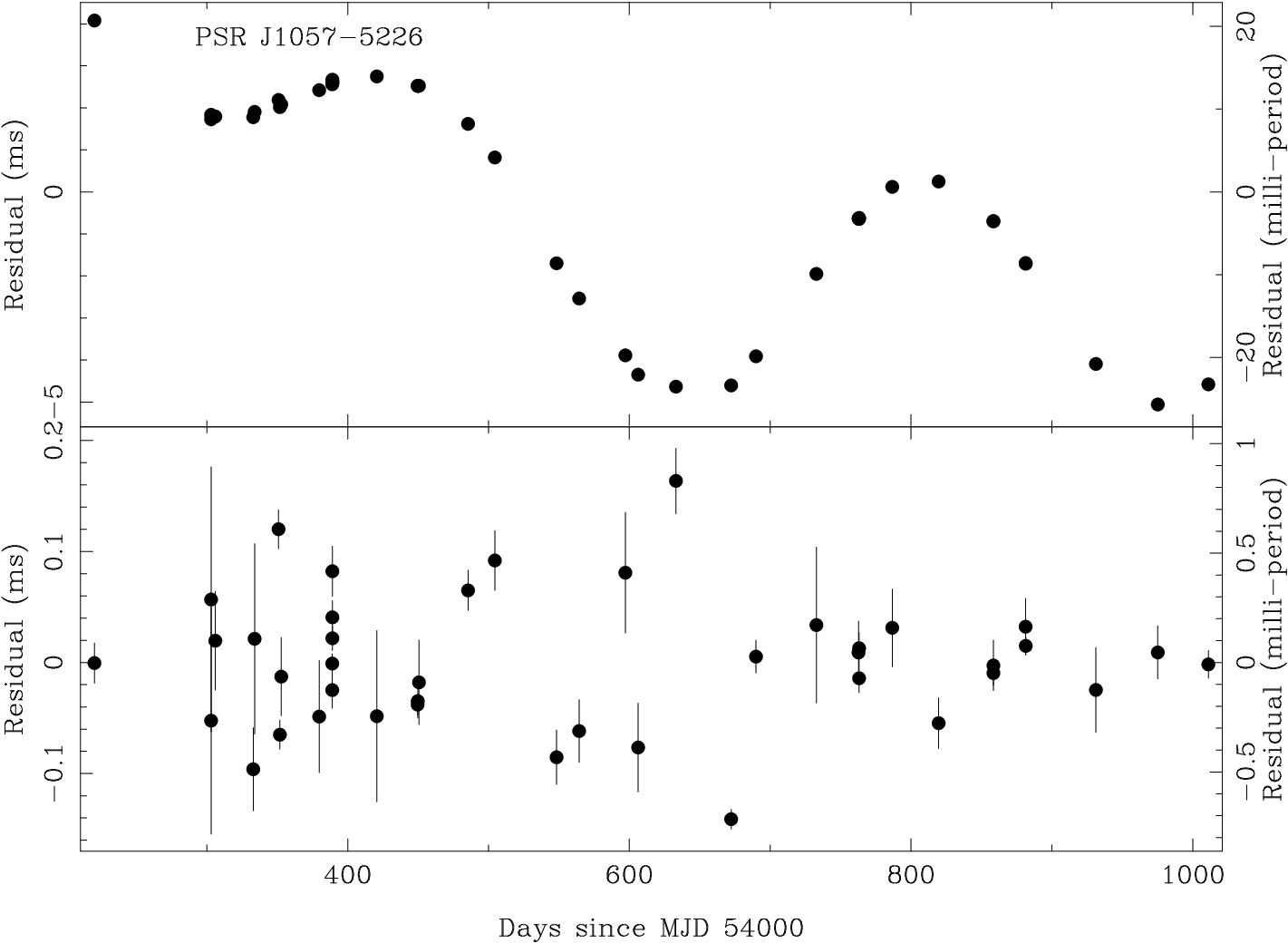} \\
\includegraphics[scale=0.3, angle=-90]{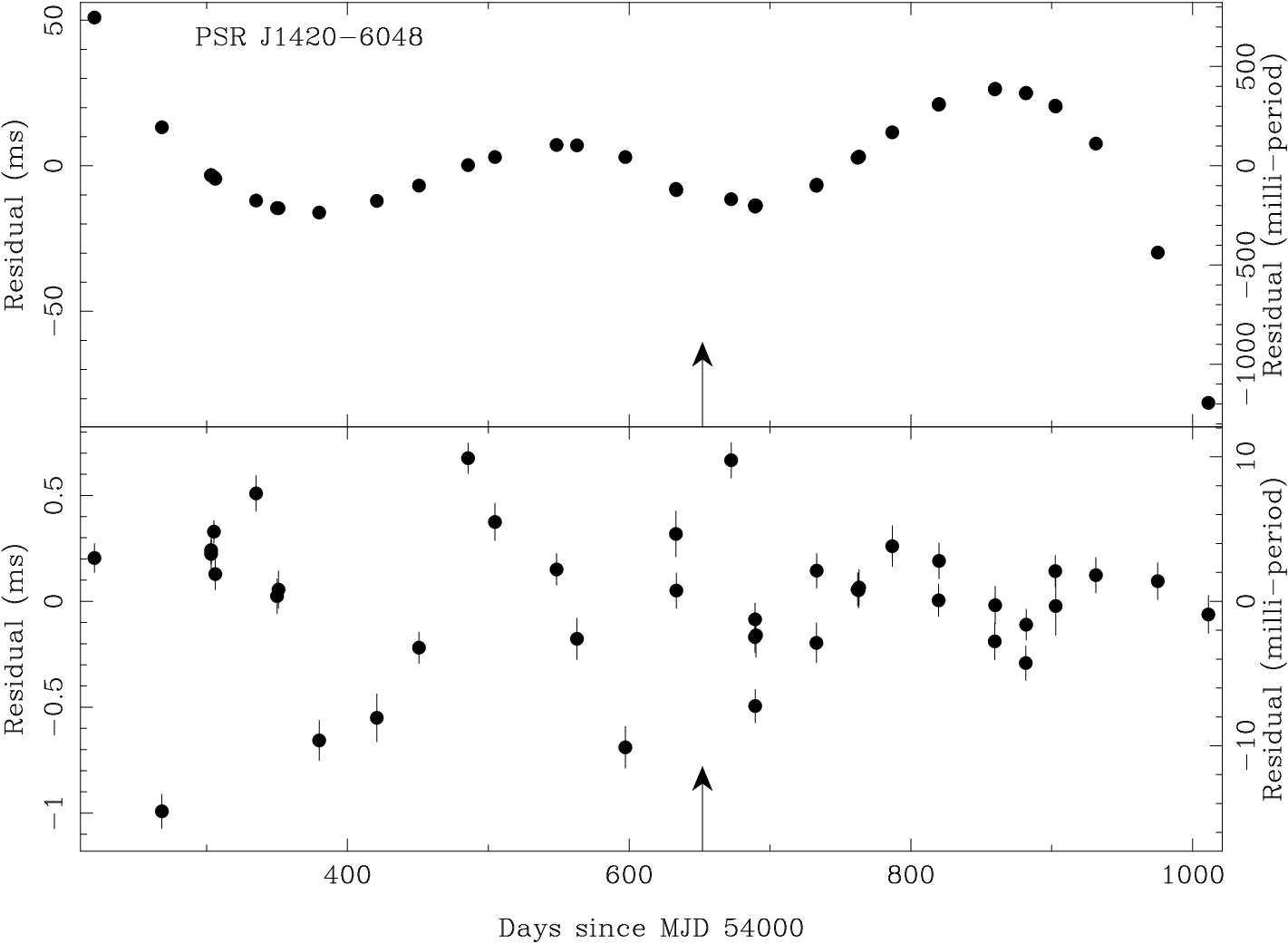} &
\includegraphics[scale=0.3, angle=-90]{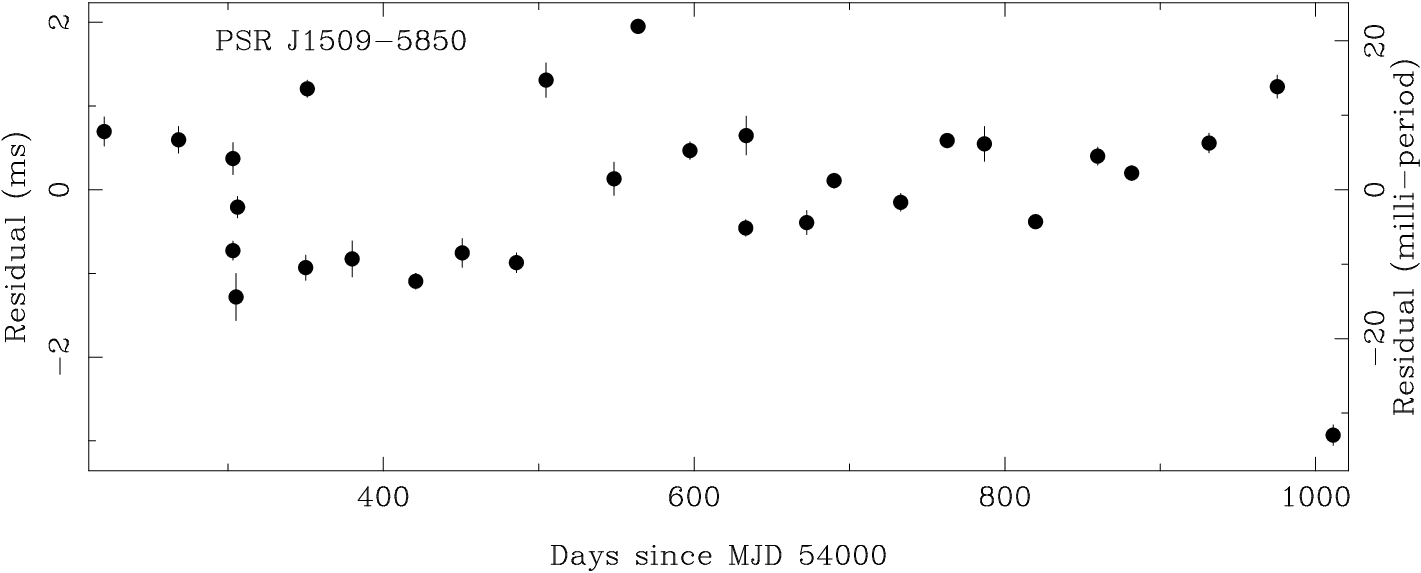} \\
\end{tabular}
\end{center}
\caption{Timing residuals for PSRs~J1048--5832 (top left),
J1057--5226 (top right), J1420--6048 (bottom left) and J1509--5850
(bottom right). Residuals before (top) and after (bottom) the use of
{\sc fitwaves} are shown for the latter three pulsars.
Note the very different scales on the y-axis for these
cases. The arrow indicates the epoch of the glitch for PSRs~J1048--5832
and J1420--6048.}
\label{figtime2}
\end{figure*}
In this section we describe in detail the ten pulsars which have been 
confirmed as gamma-ray emitters by {\it Fermi} \citep{abdo9}.
We show residuals as a function of time for each pulsar and 
provide the complete ephemerides 
in Tables~\ref{partab} and \ref{partab2}
including details of the {\sc fitwaves} parameters where
appropriate. These ephemerides, in conjunction with the publicly available
photon list from {\it Fermi}, can be used to generate gamma-ray profiles
of the {\it Fermi}-detected pulsars.

We note that for all the pulsars, the clock correction is to TT(TAI),
the solar system ephemeris used is DE405, and we are using the
barycentric dynamical time (BDT) definition for the spin parameters. 
For a full description of the meaning of these terms and their implementation
in {\sc tempo2} see Hobbs et al. (2006).

The pulsar's position, proper motion and parallax are held constant during
the fitting procedure using the best values taken from the literature.
The pulsar's DM is measured across the band at 1.4~GHz and is also held
constant during the fitting. The epoch refers to the epoch at which the
rotational frequency and positional parameters are measured.
The reference MJD and reference frequency
given in the table is the TOA of the fiducial point (the peak of the pulse
profile) at the Parkes telescope. Start and finish MJD denote
the first and last data points used and hence the range of dates over which the
fit is valid. The accuracy of the ephemerides outside the valid MJD range
deteriorates quickly in the majority of cases.

{\bf PSR~J0659+1414/B0656+14 (see Fig~\ref{figtime1}):}
This pulsar is a nearby, relatively low $\dot{E}$ pulsar.  It is known to
produce both thermal and non-thermal X-ray emission \citep{dcm+05} 
and suspected as a gamma-ray pulsator \citep{rfk+96} before confirmation
with {\it Fermi} (Weltevrede et al. 2009). Bright, ``spiky'' emission
is occasionally seen in the radio pulses \citep{wws+06}.
The pulsar has low timing
noise and is not known to have glitched in spite of decades of monitoring.
There was no need to whiten the data using {\sc fitwaves} as the timing 
noise is low.  Astrometric data are taken from \citet{btgg03}.

{\bf PSR~J0742--2822/B0740--28 (see Fig~\ref{figtime1}):}
This pulsar was not known as a high-energy emitter 
until its recent detection with {\it Fermi} (Weltevrede et al. 2009). It
is extremely bright in the radio and the rotation axis appears to point
in the direction of its proper motion \citep{jhv+05}. The pulsar is not
known to glitch but suffers a high level of timing noise which was
successfully removed with the {\sc fitwaves} algorithm.
Astrometric data are taken from \citet{fgml97}.

{\bf PSR~J0835--4510/B0833-45 (see Fig~\ref{figtime1}):}
The Vela pulsar was one of the first pulsars discovered and has a timing 
history stretching back over 40 years. Located in the Vela supernova remnant,
its gamma-ray pulsations have long been known \citep{tfko75} with
the EGRET results discussed in Kanbach et al. (1994)\nocite{kab+94} and
recent results from the Agile satellite are reported in
Pellizoni et al. (2009)\nocite{ppp+09}.  {\it Fermi} has
produced beautiful high quality data on this pulsar \citep{abdo3}.
Vela suffers glitches on a quasi-regular basis \citep{wmp+00}, but no large
glitch has occurred over the timespan of our data.
The position, proper motion and parallax for Vela are taken from \citet{dlrm03}.
\begin{figure}[t]
\begin{center}
\begin{tabular}{c}
\includegraphics[scale=0.3, angle=-90]{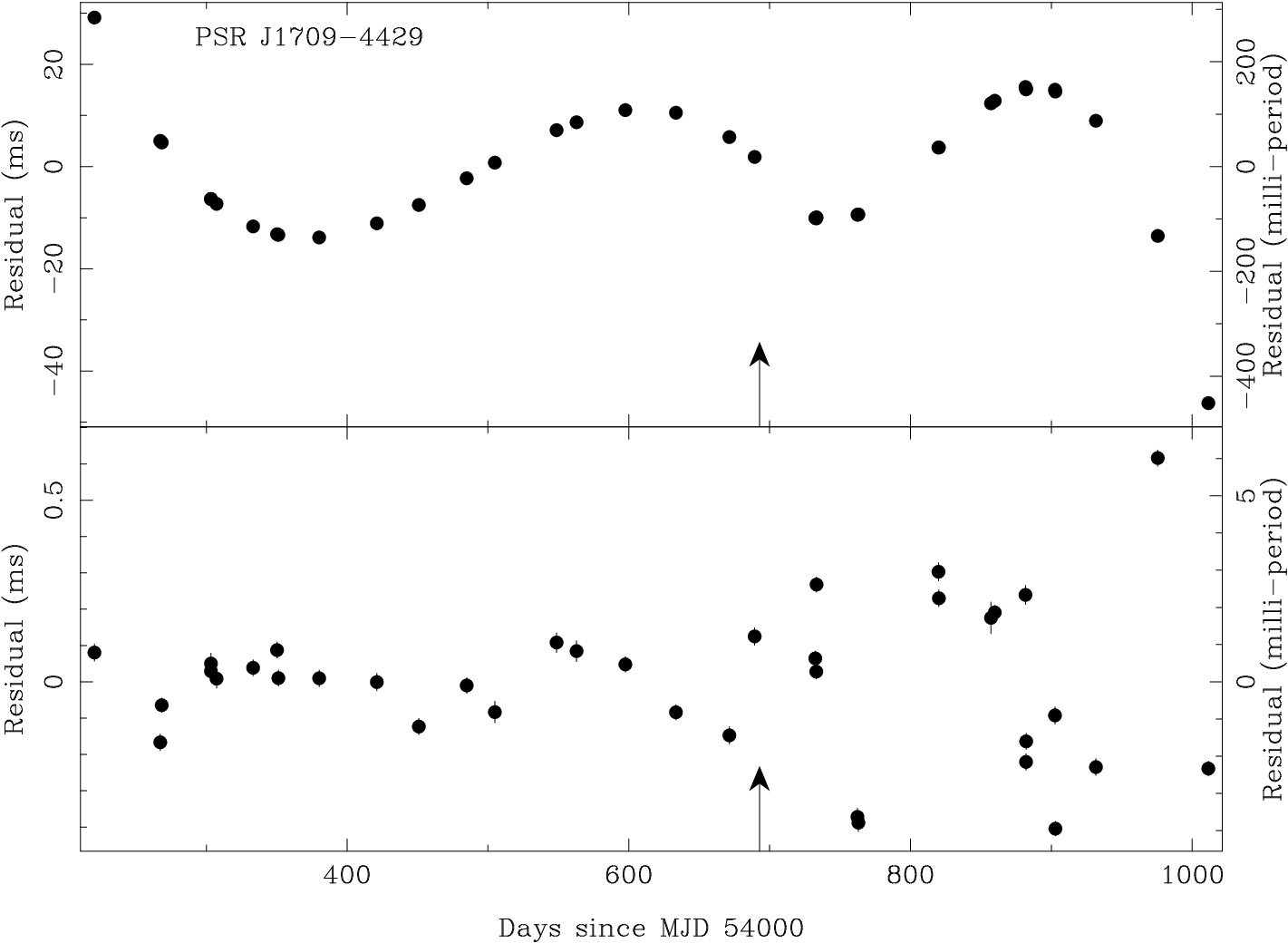} \\
\includegraphics[scale=0.3, angle=-90]{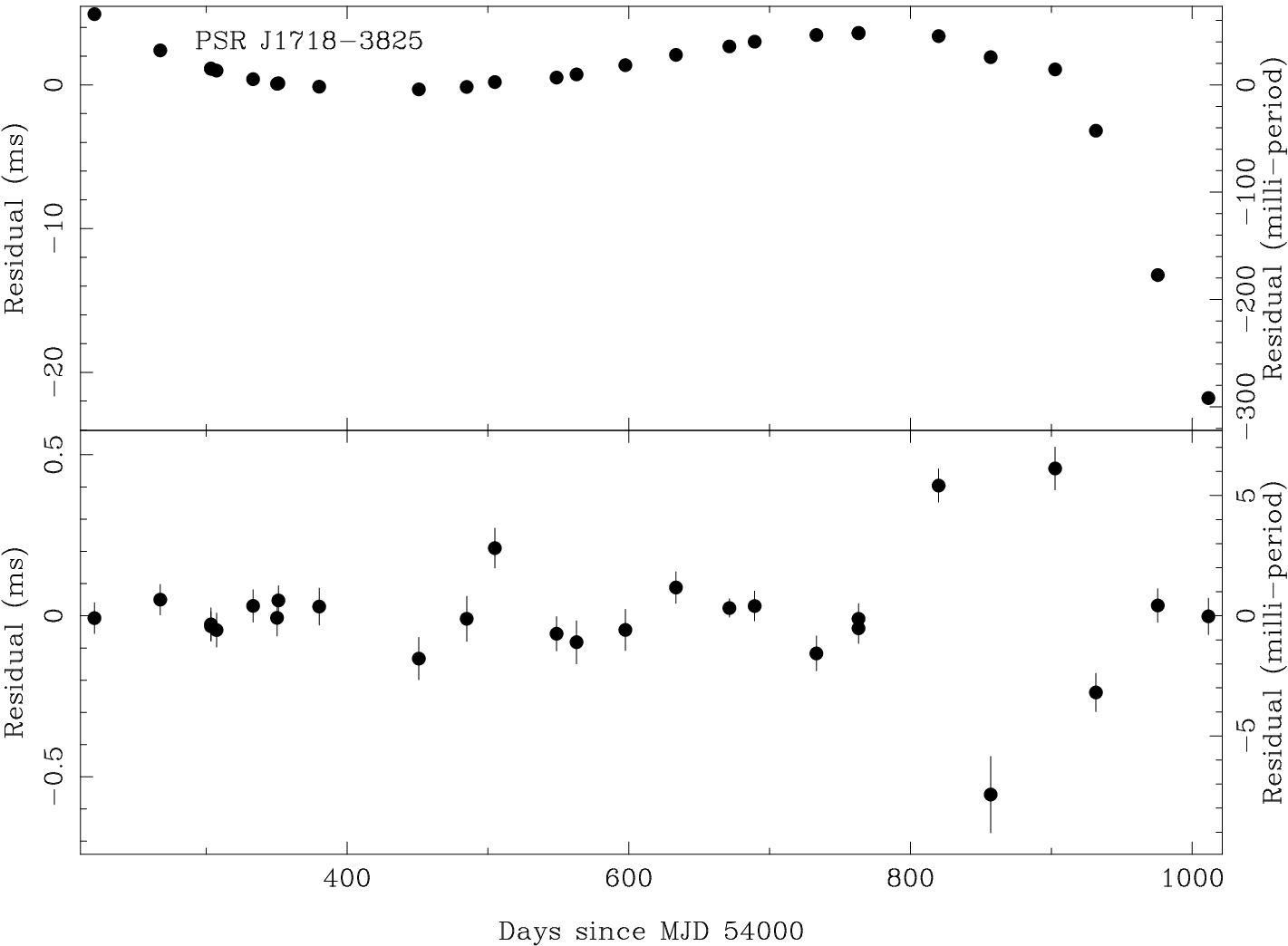} \\
\end{tabular}
\end{center}
\caption{Timing residuals for PSRs~J1709--4429 (top),
J1718--3825 (bottom). Residuals before (top) and after (bottom) the use of
{\sc fitwaves} are shown.
Note the very different scales on the y-axis. The arrow indicates the epoch 
of the glitch for PSR~J1709--4429.}
\label{figtime3}
\end{figure}

{\bf PSR~J1028--5819 (see Fig~\ref{figtime1}):}
This pulsar was discovered in 2008 March in a search
of unidentified EGRET error boxes \citep{kjk+08} and subsequently
detected as a pulsed gamma-ray emitter by {\it Fermi} \citep{abdo2}.
The very narrow radio profile allows for accurate
TOAs to be obtained and the residual is only 40$\mu$s.
The pulsar position is from the interferometric observations of \citet{kjk+08}.

{\bf PSR~J1048--5832/B1046--58 (see Fig~\ref{figtime2}):}
This pulsar is a `Vela-like' object, proposed
tentatively as a gamma-ray emitter from EGRET data \citep{klm+00} and
confirmed by {\it Fermi} \citep{abdo10}.
Large pulses are occasionally
seen on the rising edge of the profile \citep{jr02}. The pulsar glitched in
early 2008. The timing noise is large and tracks smoothly over
the glitch epoch. The pulsar's position is that from \citet{wmp+00}.

{\bf PSR~J1057--5226/B1055--52 (see Fig~\ref{figtime2}):}
This pulsar has an interpulse and both the main and
interpulses have a complex pulse morphology, described extensively in
\citet{ww09}. Although it has a relatively low $\dot{E}$,
it was seen as a pulsed gamma-ray emitter by EGRET
(Fierro et al. 1993\nocite{fbb+93}).  The gamma-ray emission is claimed to 
be associated with the pole that produces the radio interpulse
emission (e.g. Weltevrede \& Wright 2009)\nocite{ww09}.
The timing noise seems quasi-periodic with a period close to one year.
The pulsar's position is taken from the radio timing paper
of Newton et al. (1981)\nocite{nmc81}.

{\bf PSR~J1420--6048 (see Fig~\ref{figtime2}):}
This pulsar lies in a complex region of the
Galactic plane and is a known X-ray emitter \citep{rrj01}. It was mooted
as a possible counterpart to an EGRET unidentified source \citep{dkm+01},
and confirmed as a gamma-ray emitter by {\it Fermi} (Weltevrede et al. 2009).
A glitch in the pulsar happened in mid 2008, just before {\it Fermi} began
its all sky survey. The timing noise is large. The pulsar's position
is from D'Amico et al. (2001)\nocite{dkm+01}.

{\bf PSR~J1509--5850 (see Fig~\ref{figtime2}):}
The pulsar is located inside an X-ray \citep{kmp+08} and radio \citep{hb07}
wind nebula. Weltevrede et al. (2009) report the gamma-ray detection of the
pulsar.  Its timing noise is very low, the residual after fitting for 
$\nu$ and $\dot{\nu}$ is only 20 milli-periods and 
there is no need for whitening using {\sc fitwaves}. The pulsar's position
is taken from the discovery paper (Kramer et al. 2003\nocite{kbm+03}).

{\bf PSR~J1709--4429/B1706--44 (see Fig~\ref{figtime3}):}
This pulsar is a `Vela-like' object, detected as
a gamma-ray emitter by EGRET (Thompson et al. 1992)\nocite{tab+92}.
A glitch occurred in 2008 August; the daily observations and monitoring
by {\it Fermi} allows an accurate epoch to be obtained \citep{abdo11}.
The timing noise continues smoothly through the glitch. The pulsar position is 
that from \citet{wmp+00}.

{\bf PSR~J1718--3825 (see Fig~\ref{figtime3}):}
The pulsar has an associated X-ray nebula \citep{hfc+07}
and a TeV source (Aharonian et al. 2007)\nocite{aab+07}.
The gamma-ray detection of the pulsar is reported in
Weltevrede et al. (2009). Its timing noise is relatively large and has a 
longer timescale than many of the other pulsars in our sample.
The pulsar's position is taken from Manchester et al. (2001)\nocite{mlc+01}.

\section{Summary}
We started a comprehensive program of pulsar timing with the Parkes radio
telescope in 2007 February. The launch of the {\it Fermi} satellite in 2008 June
has resulted in the detection of more than 20 (non-MSP) radio pulsars at 
gamma-ray wavelengths, of which 10 are observed at Parkes.
In this paper we have provided
the timing ephemerides for the 10 pulsars, which, in conjunction with the
publicly available photon parameters from {\it Fermi}, allow a recreation
of the gamma-ray profiles.  In addition we have detected 14 glitches in 
13 different pulsars and provide the glitch parameters.

The {\it Fermi} mission will continue for at least a further 4 years and 
we plan to carry out the pulsar timing for a similar length of time. 
Researchers who require ephemerides from any of the pulsars contained in 
Table~\ref{tab1a} can contact the authors of this paper.

\section*{Acknowledgments}
We thank the many observers who have helped with the observations at Parkes
over the last two years.  The Australia Telescope is funded by
the Commonwealth of Australia for operation as a National Facility
managed by the CSIRO. BK was supported by NASA and NSF grant AST-0506453.

\bibliographystyle{mn2e}
\bibliography{journals,glitch,modrefs,psrrefs}

\begin{sidewaystable*}[t]
\begin{center}
\caption{Timing parameters for 5 pulsars detected in gamma-rays
by {\it Fermi}.}
\label{partab}
\begin{tabular}{lccccc}
\hline
Param & J0659+1414 & J0742--2822 & J0835--4510 & J1028--5819 & J1048--5832\\
\hline
Right Ascension (J2000) & 
06:59:48.134  & 
07:42:49.026  & 
08:35:20.61149  & 
10:28:27.95  & 
10:48:12.2  \\ 
Declination (J2000) & 
+14:14:21.5  & 
--28:22:43.70  & 
--45:10:34.8751  & 
--58:19:05.225  & 
--58:32:05.8  \\ 
Spin Frequency, $\nu$, (s$^{-1}$) & 
2.598136794(2) & 
5.996283159(8)  & 
11.191119(5)  & 
10.9404949958(3) & 
8.08398924(6) \\ 
$\dot{\nu}$ (10$^{-15}$) & 
--370.981(5) & 
--611(7) & 
--15600(170) & 
--1928.4(3) & 
--6259(7) \\ 
DM (cm$^{-3}$pc) & 
13.7(2)  & 
73.790(3) & 
67.9754(8)  & 
96.39(9)  & 
128.822(8)  \\ 
Proper Motion in RA (mas/yr)& 
44.07  & 
--29  & 
--49.68  & 
 & 
 \\ 
Proper Motion in Dec (mas/yr) & 
--2.4  & 
4  & 
29.9  & 
 & 
 \\ 
Epoch (MJD) & 
49721.0  & 
54615.0  & 
54217.0  & 
54787.0  & 
54615.0  \\ 
Ref. MJD & 
54762.6946809825  & 
54632.9827544637  & 
54606.2500568554  & 
54786.6818285185  & 
54606.2571751136  \\ 
Ref. Freq (MHz) & 
1369.235  & 
1377.652  & 
1373.864  & 
1374.656  & 
1373.864  \\ 
Glitch Epoch (MJD) &
&
&
&
&
54490 \\
Glitch Phase Offset &
&
&
&
&
--0.484(6) \\
Glitch $\Delta\nu$ (10$^{-5}$) &
&
&
&
&
2.4616(3) \\
Glitch $\Delta\dot{\nu}$ (10$^{-14}$) &
&
&
&
&
--3.7(4) \\
Start (MJD) & 
54505  & 
54220  & 
54207  & 
54563  & 
54220  \\ 
Finish (MJD) & 
55011  & 
55010  & 
55010  & 
55011  & 
55011  \\ 
Wave Frequency (rad/day) & 
 & 
0.0063560           & 
0.0052116           & 
0.0093621           & 
0.0059587           \\ 
Wave 1 & 
 & 
--0.4498,+0.05777  & 
+3.318,--1.538  & 
--0.008672,--0.002189 & 
--0.3531,--0.1548  \\ 
Wave 2 & 
 & 
+0.09547,--0.02979  & 
+0.1863,--1.021  & 
+0.002582,--0.001674  & 
+0.07018,+0.01160  \\ 
Wave 3 & 
 & 
--0.03240,+0.0167 & 
--0.3375,--0.2816  & 
--0.0002706,--3.2135e-05  & 
--0.01741,--0.004765  \\ 
Wave 4 & 
 & 
+0.01348,--0.009575   & 
--0.1746,+0.08070 & 
 & 
+0.007688,+0.0005985  \\ 
Wave 5 & 
 & 
--0.004789,+0.005083  & 
+0.003159,+0.07619   & 
 & 
--0.001823,+0.002077  \\ 
Wave 6 & 
 & 
+0.001260,--0.002011  & 
+0.02241,+0.007917  & 
 & 
 \\ 
Wave 7 & 
 & 
--0.0003854,+0.0008335  & 
+0.003275,--0.003641  & 
 & 
 \\ 
\hline
\end{tabular}
\end{center}
\end{sidewaystable*}

\begin{sidewaystable*}[t]
%\addtocounter{table}{-1}
\begin{center}
\caption{Timing parameters for 5 pulsars detected in gamma-rays
by {\it Fermi}.}
\label{partab2}
\begin{tabular}{lccccc}
\hline
Param & J1057--5226 & J1420--6048 & J1509--5850 & J1709--4429 & J1718--3825\\
\hline
Right Ascension (J2000) & 
10:57:58.84  & 
14:20:08.237  & 
15:09:27.13  & 
17:09:42.728  & 
17:18:13.565  \\ 
Declination (J2000) & 
--52:26:56.3  & 
--60:48:16.43  & 
--58:50:56.1  & 
--44:29:08.24  & 
--38:25:18.06  \\ 
Spin Frequency, $\nu$, (s$^{-1}$) & 
5.0732274494(7) & 
14.66248295(6)  & 
11.2455254698(1) & 
9.756518994(9)  & 
13.391571381(4) \\ 
$\dot{\nu}$ (10$^{-15}$) & 
--150.4(6)  & 
--17710(16)  & 
--1159.23(1) & 
--8831.0(7)  & 
--2375(4)  \\ 
Epoch (MJD) & 
54615.0  & 
54615.0  & 
54615.0  & 
54616.0  & 
54616.0  \\ 
DM (cm$^{-3}$pc) & 
30.1  & 
360  & 
137.7  & 
75.69  & 
247.4  \\ 
Ref. MJD & 
54606.2654505538  & 
54633.2224514535  & 
54633.2560188871  & 
54633.4457155762  & 
54633.4498848905  \\ 
Ref. Freq (MHz) & 
1373.864  & 
1377.652  & 
1377.652  & 
1377.652  & 
1377.652  \\ 
Glitch Epoch (MJD) &
&
& 54660
&
& 54692.8
\\
Glitch Phase Offset &
&
& --0.23(2) 
&
& --0.178(9)
\\
Glitch $\Delta\nu$ (10$^{-5}$) &
&
& 1.3659(4)
&
& 2.6837(3)
\\
Glitch $\Delta\dot{\nu}$ (10$^{-14}$) &
&
& --10.0(3)
&
& --6.2(1)
\\
Start (MJD) & 
54220  & 
54220 & 
54220  & 
54220  & 
54220  \\ 
Finish (MJD) & 
55011  & 
55011  & 
55011  & 
55011  & 
55011  \\ 
Wave Frequency (rad/day) & 
0.0059587  & 
0.0039726  & 
 & 
0.0061797  & 
0.0059593  \\ 
Wave 1 & 
--0.02318,--0.0007458  & 
+5.079,--1.557   & 
 & 
+0.06933,--0.2996   & 
--0.3333,+0.007675   \\ 
Wave 2 & 
+0.005196,+0.0001067  & 
--0.6315,+0.04927   & 
 & 
+0.01173,+0.0304 & 
+0.06940,--0.009321  \\ 
Wave 3 & 
+0.0009044 ,+0.0004835  & 
+0.08892,--0.007211  & 
 & 
--0.002489,--0.001332   & 
--0.02062,+0.006114  \\ 
Wave 4 & 
+0.0003618,+0.0003014  & 
 & 
 & 
 & 
+0.005628,--0.003100  \\ 
Wave 5 & 
--0.0001134,--2.6808e-05  & 
 & 
 & 
 & 
--0.0009510,+0.0008923  \\ 
\hline
\end{tabular}
\end{center}
\end{sidewaystable*}

\end{document}